\documentclass{aa}  

\usepackage{graphicx}
\usepackage{txfonts}
\usepackage{tabularx}
\usepackage{inputenc}
\usepackage{threeparttable}
\usepackage{tablefootnote}
\usepackage{natbib}
\bibpunct{(}{)}{;}{a}{}{,}
\usepackage{url}
\newcommand{\fermi}{\textit{Fermi}}  

\newcommand\eflux{\mbox{${\rm \, erg \,\, cm^{-2} \, s^{-1}}$}}

\begin{document}

   \title{Exploring the connection between radio and GeV-TeV $\gamma$-ray emission in the 1FHL and 2FHL AGN samples}

   \author{R.~Lico\inst{1,2}\fnmsep\thanks{Email: rocco.lico@unibo.it}, M.~Giroletti\inst{1}, M.~Orienti\inst{1,2}, L.\ Costamante\inst{3}, V.~Pavlidou\inst{4}, F.~D'Ammando\inst{1,2}, F.~Tavecchio\inst{5}.} 

   \institute{INAF - Istituto di Radioastronomia, via Gobetti 101, 40129 Bologna, Italy.
\and Dipartimento di Fisica e Astronomia, Universit\`a di Bologna, via Gobetti 93/3, 40129 Bologna, Italy.
\and ASI – Unit\`a Ricerca Scientifica, Via del Politecnico snc, I-00133, Roma, Italy.
\and Department of Physics and Institute for Plasma Physics, University of Crete, 71003 Herak-lion, Greece.
\and INAF – Osservatorio Astronomico di Brera, via E. Bianchi 46, I-23807 Merate, Italy.
              }
 
\date{Received ; accepted }

  \abstract
{With the advent of the \fermi\ Large Area Telescope (LAT) it was revealed that blazars, representing the most extreme radio-loud active galactic nuclei (AGN) population, dominate the census of the $\gamma$-ray sky, and a significant correlation was found between radio and $\gamma$-ray emission in the 0.1 - 100\,GeV energy range. However, the possible connection between radio and very high energy (VHE, $E > 0.1$\,TeV) emission still remains elusive, owing to the lack of a homogeneous coverage of the VHE sky.
}
{The main goal of this work is to quantify and assess the significance of a possible connection between the radio emission on parsec scale measured by the very long baseline interferometry (VLBI) and GeV-TeV $\gamma$-ray emission in blazars, which is a central issue for understanding the blazar physics and the emission processes in these objects.}
{We investigate the radio VLBI and high energy $\gamma$-ray emission by using two large and unbiased AGN samples extracted from the first and second \fermi -LAT catalogs of hard $\gamma$-ray sources detected above 10\,GeV (1FHL) and 50\,GeV (2FHL). For comparison, we perform the same correlation analysis by using the 0.1 - 300\,GeV $\gamma$-ray energy flux provided by the third \fermi -LAT source catalog (3FGL). We assess the correlation statistical significance by means of a method based on permutations of the luminosities, by taking into account the various observational biases, which may apparently enhance or spoil any intrinsic correlation.
}
{We find that the correlation strength and significance depend on the $\gamma$-ray energy range with a different behavior among the blazar sub-classes. 
Overall, the radio and $\gamma$-ray emission above 10\,GeV turns out to be uncorrelated for the full samples and for all of the blazar sub-classes with the exception of high synchrotron peaked (HSP) objects, which show a strong and significant correlation. On the contrary, when 0.1 - 300\,GeV $\gamma$-ray energies are considered, a strong and significant correlation is found for the full blazar sample as well as for all of the blazar sub-classes.
}
{We interpret and explain this correlation behavior within the framework of the blazar spectral energy distribution properties. In the most powerful blazars, in general of low synchrotron peaked type, the high energy emission component peaks at energies lower that those sampled by the LAT. On the contrary, in HSP blazars the part of the high energy spectrum affected by cooling effects is well beyond the energy range sampled by the LAT, showing a rising spectrum both in the 3FGL and 1FHL/2FHL energy ranges.}


   \authorrunning{R.\ Lico et al. 2017}
\titlerunning{Radio and GeV-TeV $\gamma$-ray emission connection in the 1FHL/2FHL AGN samples}
   \maketitle
%

\section{Introduction}
Blazars are the most extreme objects in the class of active galactic nuclei (AGNs). They can be divided into flat spectrum radio quasars (FSRQs) and BL Lac objects (BL Lacs) based on the presence or not of broad emission lines in their optical spectra \citep[e.g.\,,][]{Stickel1991}.
Their spectral energy distribution (SED) is typically dominated by two non-thermal components, extending from the radio band to $\gamma$ rays.  
The low-frequency component is due to synchrotron emission by relativistic electrons within the jet, and its peak ($\nu^{\mathrm{Syn}}_{\mathrm{peak}}$) can be found in the spectral region extending from radio to soft X-ray energies.
Depending on the $\nu^{\mathrm{Syn}}_{\mathrm{peak}}$ position, blazars are further classified as: low synchrotron peaked (LSP; $\nu^{\mathrm{Syn}}_{\mathrm{peak}} < 10^{14}$\,Hz), intermediate synchrotron peaked (ISP; $10^{14}$\,Hz $\nu^{\mathrm{Syn}}_{\mathrm{peak}} < 10^{15}$\,Hz), and high synchrotron peaked (HSP; $\nu^{\mathrm{Syn}}_{\mathrm{peak}} > 10^{15}$\,Hz) \citep{Abdo2010a}. In general FSRQs are predominantly LSP objects, while BL Lacs can be part of all of the three classes.

The high-frequency component of the SED, peaking from X-ray to $\gamma$-ray bands, is commonly assumed to originate from inverse Compton (IC) scattering of low energy photons by relativistic electrons in the jet.
The scattered photons may be the same photons produced by the synchrotron mechanism \citep[synchrotron-self-Compton, SSC, e.g.\,,][]{Maraschi1992, Abdo2010a, Hovatta2014}, or photons from external sources such as the accretion disk, the broad line region and/or the dusty torus \citep[external Compton, EC, e.g.\,,][]{Ghisellini1996, Sikora2008}.
Hadronic models, in which relativistic protons within the jet are the ultimate responsible for the observed emission, have been proposed \citep[e.g.\,,][]{Levinson2006, Bottcher2007}.

In the last years, the large area telescope (LAT) onboard the \fermi\ satellite confirmed that blazars dominate the census of the $\gamma$-ray sky \citep{Acero2015}.
Exploring the possible correlation between radio and $\gamma$-ray emission is a fundamental step to understand the physics and the emission processes in blazars, and this topic was the subject of several works \citep[e.g.\,,][]{Kovalev2009, Ghirlanda2010, Giroletti2010, Mahony2010, Nieppola2011, Piner2014, Giroletti2016}. 

\citet{Ackermann2011} revealed a positive and highly significant correlation between radio and $\gamma$-ray emission in the energy range between 100\,MeV and 100\,GeV for the AGNs included in the \fermi -LAT first source catalog \citep[1FGL, ][]{Abdo2010b}. In that work, the authors made use of both archival interferometric 8\,GHz data and concurrent single dish 15\,GHz observations from the Owens Valley Radio Observatory observing program \citep{Richards2011}.
In particular, \citet{Ackermann2011} found that the correlation strength decreases when higher $\gamma$-ray energies are considered. A similar result is reported in a more recent work by \citet{Mufakharov2015}, in which the authors explore the correlation between radio at cm wavelengths and $\gamma$-ray emission at $E > 100$\,MeV for 123 1FGL \fermi\ blazars. In that work, based on quasi-simultaneous observations, the authors find a positive and statistically significant correlation between the emission bands, that weakens when higher $\gamma$-ray energies are used.

The possible correlation between radio and very-high energy (VHE, $E > 0.1$\,TeV) $\gamma$ rays still remains, mainly due to the lack of a homogeneous coverage of the VHE sky.
Currently, VHE observations of blazars are conducted by imaging atmospheric Cherenkov telescopes (IACTs), which mainly operate in pointing mode with a limited sky coverage, and usually observe sources in flaring state. 
All of these limitations, which introduce a strong bias in VHE catalogs and make it difficult to assess any possible radio-VHE correlation, will be overcome by the advent of the new generation Cherenkov Telescope Array \citep[CTA, ][]{Actis2011}.

\begin{table}
\caption{\small Composition of the two source sample extracted from 1FHL and 2FHL catalogs.}
\label{tab_sample_composition}    
\centering   
\small            
\setlength{\tabcolsep}{9pt}
\renewcommand{\arraystretch}{1.3}
\begin{tabular}{lccc} 
\hline
\hline 
Source type & Catalog & Num. of Sources & Sources with $z$ \\
\hline
\hline           
All sources & 1FHL & 237 & 147   \\
            & 2FHL & 131 &  76   \\
BL Lac      & 1FHL & 173 & 100   \\
            & 2FHL & 112 &  63   \\
FSRQ        & 1FHL &  44 &  44   \\ 
            & 2FHL &   5 &   5   \\
HSP         & 1FHL & 103 &  60   \\ 
            & 2FHL &  84 &  48   \\
ISP         & 1FHL &  45 &  23   \\
            & 2FHL &  18 &   7   \\ 
LSP         & 1FHL &  58 &  52   \\
            & 2FHL &  23 &  17   \\
\hline
\end{tabular}
\end{table}

Among the 61 currently known TeV blazars included in the online TeVCat\footnote{\url{http://tevcat.uchicago.edu/}} catalog, which contains all of the TeV sources so far detected, 75\% (46 objects) of them belong to the HSP class\footnote{We refer to the catalog version 3.400.}. 
In general, HSP blazars show peculiar features such as lower Compton dominance, lower synchrotron luminosity, and parsec scale jets less variable in flux density and structure than in other blazars \citep[e.g.\,,][]{Giroletti2004, Piner2008, Lico2012, Blasi2013, Lico2014}. 

At present, the first and second \fermi -LAT catalogs of high-energy $\gamma$-ray sources, 1FHL and 2FHL \citep[][]{Ackermann2013, Ackermann2016}, represent the best compromise for addressing the connection between radio and hard $\gamma$-ray emission. The 1FHL (in the 10 - 500\,GeV energy range) and the 2FHL (in the 50\,GeV - 2\,TeV energy range) catalogs provide us with two large, deep and unbiased samples of $\gamma$-ray sources in an energy range approaching and partly overlapping with the VHE band.

In this work we investigate the possible radio-VHE correlation by performing a statistical analysis on the 1FHL and 2FHL AGN samples, mostly composed of HSP blazars, by using the method developed by \citet{Pavlidou2012}. 
A preliminary version of the third Fermi-LAT catalog of high-energy gamma-ray sources (3FHL \footnote{Preliminary 3FHL release: arXiv:1702.00664}) has been recently released by the \fermi -LAT Collaboration, but it is not yet published. For this reason we do not use the 3FHL sources in the present analysis, and the 3FHL will be the subject of a future work.
At radio frequencies we make use of very long baseline interferometry (VLBI) flux densities, which are representative of the emission from the innermost (milliarcsecond) source region, where the $\gamma$-ray emission is likely produced.

The paper is organized as follows: in Sect.~\ref{sec.catalogs} we describe the catalogs used in this work and the sample construction; we present the results in Sect.~\ref{sec.results} and we discuss them in Sect.~\ref{sec.discussion}. Throughout the paper we use a $\Lambda$CDM cosmology with $h = 0.71$, $\Omega_m = 0.27$, and $\Omega_\Lambda=0.73$ \citep{Komatsu2011}.

\begin{figure}
\begin{center}
\includegraphics[bb=20 4 490 340, width= 0.5\textwidth, clip]{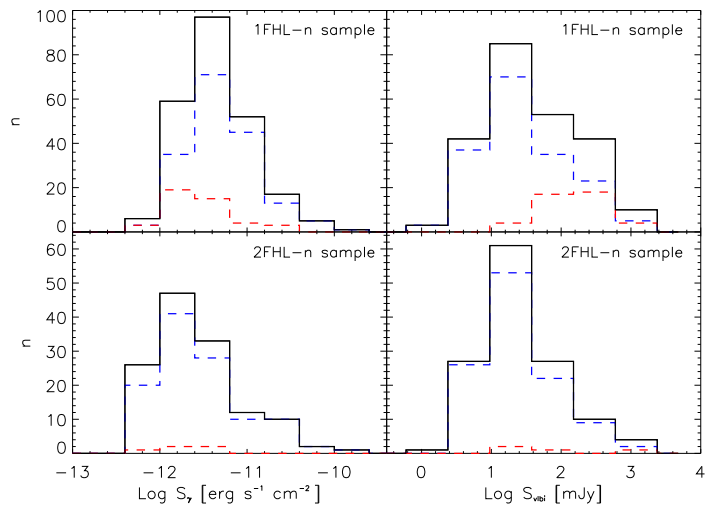} \\ 
\end{center}
\caption{\small $\gamma$-ray energy flux distribution for the 1FHL-n (top-left panel) and 2FHL-n (bottom-left panel) samples. VLBI flux density distribution for the 1FHL-n (top-right panel) and 2FHL-n (bottom-right panel) samples. The black solid lines represent the full source samples, while the blue and red dashed lines represent BL Lacs and FSRQs, respectively.}
\label{histo_flux}
\end{figure}

\begin{figure*}
\begin{center}
\includegraphics[bb=6 0 495 350, width= 0.45\textwidth, clip]{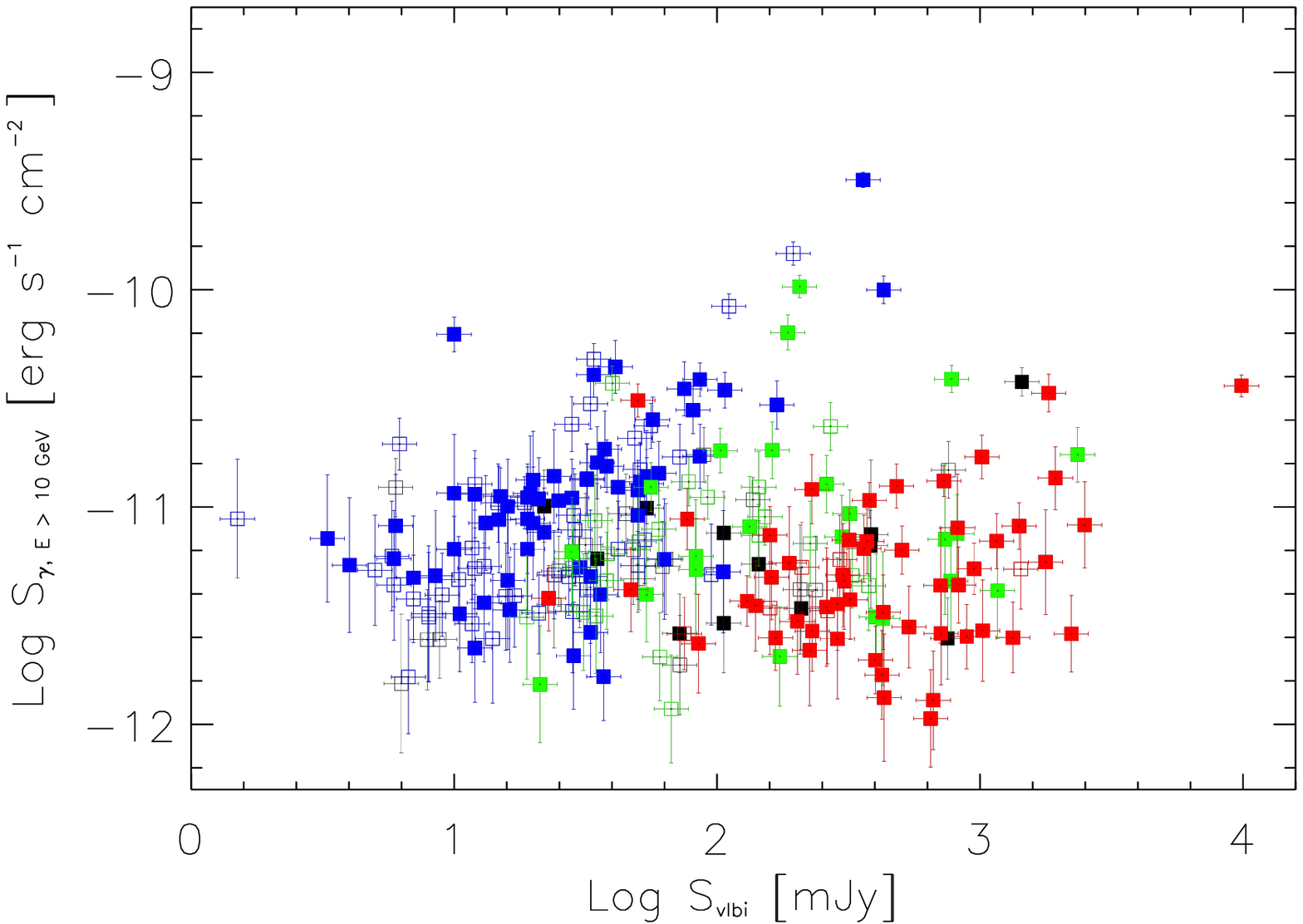} 
\includegraphics[bb=0 0 485 350, width= 0.45\textwidth, clip]{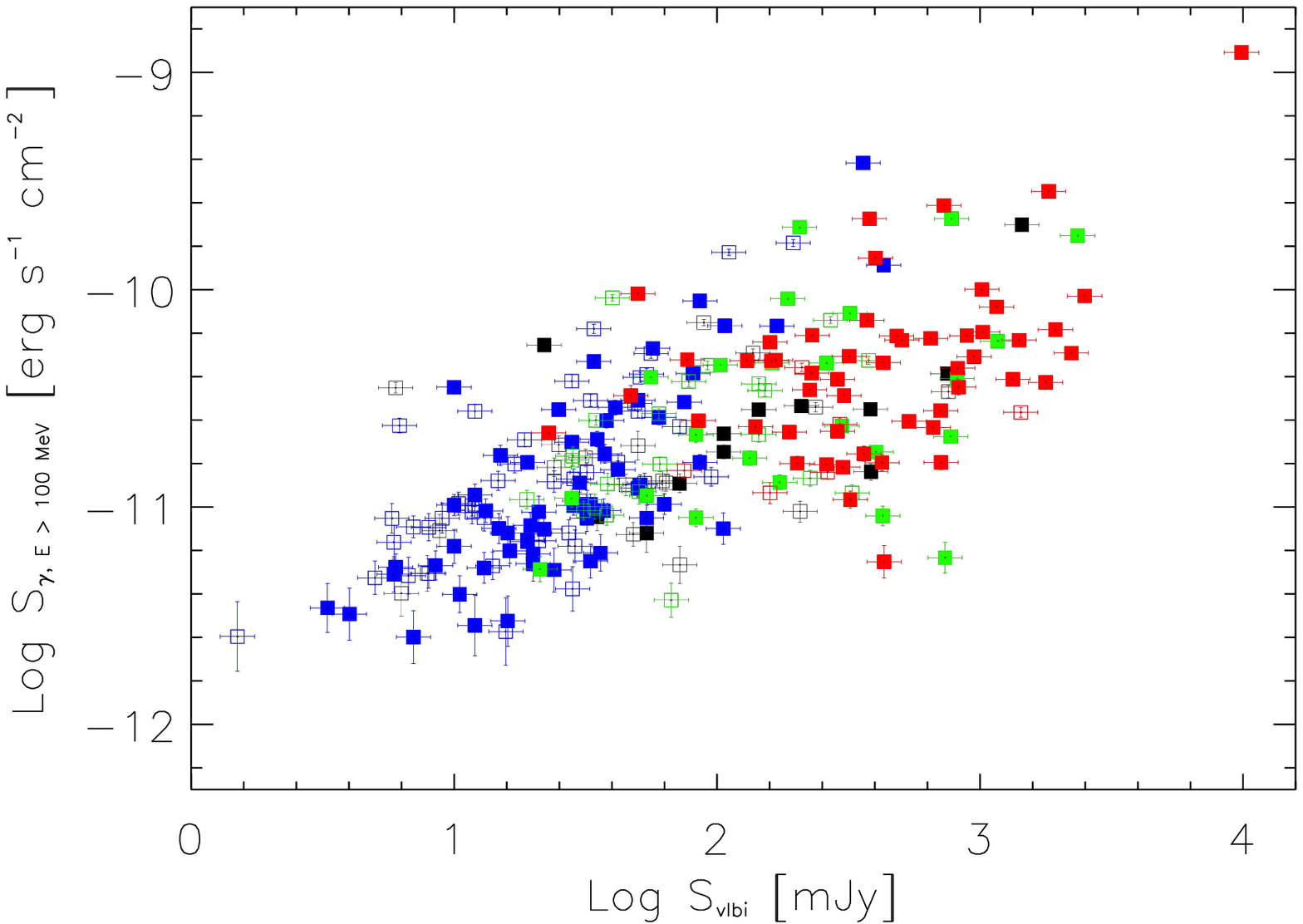}  \\
\end{center}
\caption{\small VLBI flux density vs. 1FHL (left panel) and 3FGL (right panel) energy flux scatter plots for the full 1FHL-n sample. HSP, ISP and LSP sub-classes are indicated in blue, green and red colors, respectively. Sources with no spectral classification are indicated in black color. The filled and empty symbols represent sources with or without redshift, respectively.}
\label{1fhl_scatter_plot_combo}
\end{figure*}

\begin{figure*}
\begin{center} 
\includegraphics[bb=5 155 490 355, clip]{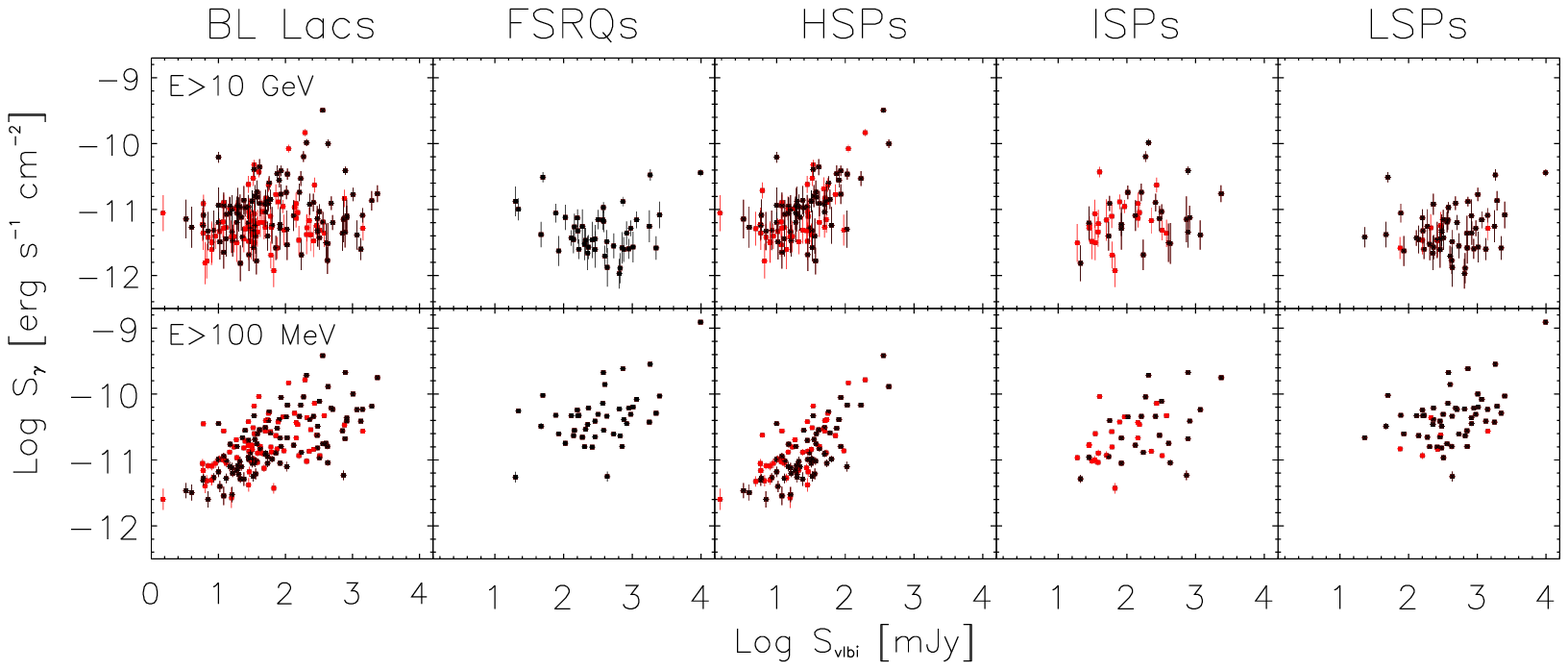} 
\end{center}
\caption{\small VLBI flux density vs. 1FHL (upper panels) and 3FGL (lower panels) energy flux scatter plots for BL Lacs, FSRQs, HSPs, ISPs,and LSPs, belonging to the 1FHL-n sample. The black and red symbols represent sources with or without redshift, respectively.}
\label{1fhl_scatter_plots}
\end{figure*}

\begin{table*}
\caption{\small Results of the correlation analysis between 1FHL (10 - 500\,GeV) energy fluxes and VLBI flux densities for various 1FHL-n sub-samples. For comparison the same analysis was performed by using 3FGL (0.1 - 300\,GeV) energy fluxes.}
\label{tab_1fhl_corr}     
\centering    
\small             
\setlength{\tabcolsep}{9pt}
\renewcommand{\arraystretch}{1.3}
\begin{tabular}{lccccc}  
\hline
\hline  
Source type & Catalog & Number of Sources & Number of $z$-bins & r-Pearson & Significance \\ 
\hline            
All sources & 1FHL & 147 & 14 & -0.05 & $ 0.59 $              \\
            & 3FGL & 147 & 14 &  0.71 & $ <10^{-6} $          \\
BL Lac      & 1FHL & 100 & 9  &  0.12 & $ 0.55 $              \\ 
            & 3FGL & 100 & 9  &  0.70 & $ <10^{-6} $          \\ 
FSRQ        & 1FHL & 44  & 4  & -0.01 & $ 0.99 $              \\  
            & 3FGL & 44  & 4  &  0.49 & $ <10^{-6} $          \\
HSP         & 1FHL & 60  & 5  &  0.57 & $ 1.0\times 10^{-6} $ \\  
            & 3FGL & 60  & 5  &  0.77 & $ <10^{-6} $          \\
ISP         & 1FHL & 23  & 2  &  0.19 & $ 0.40 $              \\ 
            & 3FGL & 23  & 2  &  0.46 & $ 2.5\times 10^{-2} $ \\  
LSP         & 1FHL & 52  & 5  &  0.21 & $ 0.12 $              \\ 
            & 3FGL & 52  & 5  &  0.43 & $ 3.0\times 10^{-6} $ \\ 
\hline
\end{tabular}
\end{table*}

\section{Catalogs and sample selection}
\label{sec.catalogs}
\subsection{1FHL}
The 1FHL catalog is based on LAT data accumulated during the first 3 years of \fermi\ operation (from 2008 August to 2011 August), providing us with a uniform and deep all-sky survey in the 10 - 500\,GeV energy range.
The 1FHL contains 514 $\gamma$-ray sources detected with Test Statistic (TS) larger than 25 (significance above $\sim 4\sigma$), and provides for each source the position (with a mean $95\%$  positional  confidence radius of $\sim5.3$ arcmin), the spectral and variability properties, as well as the associations with sources at other wavelengths.
65 ($\sim$13\%) out of the 1FHL sources do not have any plausible low-frequency association and are classified as unassociated $\gamma$-ray sources (UGS).
Among the 449 associated 1FHL sources, 393 are AGNs while the remaining ones (12\%) are sources of Galactic nature (i.e. pulsars, supernova remnants, and pulsar wind nebulae).
We note that 88\% of the 1FHL associated sources are statistically associated with blazars (the 75\% of the entire 1FHL catalog), which clearly indicates that the LAT sky at energies $>10$\,GeV is dominated by blazars.

\subsection{2FHL}
The 2FHL catalog is based on data accumulated during the first 6.5 years of the \fermi\ mission, from 2008 August to 2015 April, at the highest LAT energy range between 50\,GeV and 2\,TeV.
The 2FHL contains 360 $\gamma$-ray sources detected above 4$\sigma$ significance and represents the largest, deepest and unbiased sample of $\gamma$-ray sources in the VHE domain: about 80\% (284/360) of the 2FHL sources have photons detected at $E > 100$\,GeV.
For each source the 2FHL catalog provides: the position (with a mean positional confidence radius of $\sim4.0$ arcmin at 95\% confidence level), the spectral and variability properties, and the possible multi-frequency association.
The vast majority of the 2FHL sources are AGNs (76\%), and 98\% of them are statistically associated with blazars. 
Of the remaining 2FHL sources, 11\% are of Galactic nature, while 13\% (48 sources) are UGS or associated with a TeV source of unknown type.

\subsection{3FGL}
The third \fermi -LAT source catalog \citep[3FGL, ][]{Acero2015} is based on LAT data accumulated during the first 4 years of the mission (from 2008 August to 2012 July). The 3FGL contains 3033 $\gamma$-ray sources detected above 4$\sigma$ significance at energies between 100\,MeV and 300\,GeV, and represents the deepest catalog in this energy range.
About $35\%$ of the 3FGL sources have no clear counterpart at low frequencies.
For each source the 3FGL catalog provides the source location region (with a mean $95\%$  positional  confidence  radius  of $\sim6.2$ arcmin), the flux measurements in different energy bins, the spectral properties, and the multi-wavelength associations.

\subsection{Radio fundamental catalog}
The Radio Fundamental Catalog\footnote{\url{http://astrogeo.org/rfc/}} (RFC) collects and provides archival VLBI flux densities and precise positions (accuracy at milliarcsecond level), at several frequencies (between 2 and 22\,GHz), for thousands of compact radio sources. 
The RFC makes use of all the available VLBI observations obtained during the past 35 years under absolute astrometry and geodesy programs.
The last RFC available release (rfc\_2016c), used in this work, is updated at 2016/07/18 and contains $11448$ objects.

\subsection{Sample selection and construction}
By considering the high Galactic latitude (|$b$|>10$^{\circ}$) AGN distribution in the sky, we notice that $68$\% (i.e.\,165/243) of 1FHL BL Lacs and $64$\% (i.e.\,43/67) of 1FHL FSRQs are found in the northern hemisphere, while $75$\% (i.e. 30/40) of AGNs of unknown type are found in the southern hemisphere. A similar fraction is found for the 2FHL AGNs.
This asymmetry in the source count distribution has not a physical origin, but it is rather due to a more sparse optical coverage in the southern hemisphere, that prevents an accurate source association. 
To avoid any possible bias introduced by the source distribution asymmetry due to the lack of a spectroscopic classification, we focus our attention on the 1FHL and 2FHL AGNs with declination $\delta > 0^{\circ}$. 

Due to the large source positional uncertainty of the $\gamma$-ray sources we make use of the coordinates of the proposed low-energy counterparts as listed in the 1FHL/2FHL catalogs. 
To obtain high resolution radio observations we cross-match our sample with the RFC. 
Being the available RFC 5\,GHz VLBI flux densities consistent with those at 8\,GHz (their average spectral index is $0.0 \pm 0.1$), we use either 5\,GHz or 8\,GHz RFC flux densities for our analysis.
For those sources not included in the RFC we use the 5\,GHz very long baseline array flux densities reported in \citet{Lico2016}.

The resulting samples that we use for the correlation analysis contain 237 1FHL sources (hereafter 1FHL-n) and 131 2FHL sources (hereafter 2FHL-n). For some sources of the selected samples either the optical (BL Lacs and FSRQs) or spectral (HSPs, ISPs and LSPs) classification is not available. The details and the composition of both samples are reported in Table~\ref{tab_sample_composition}. 


\begin{figure*}
\begin{center}
\includegraphics[bb=6 0 495 348, width= 0.45\textwidth, clip]{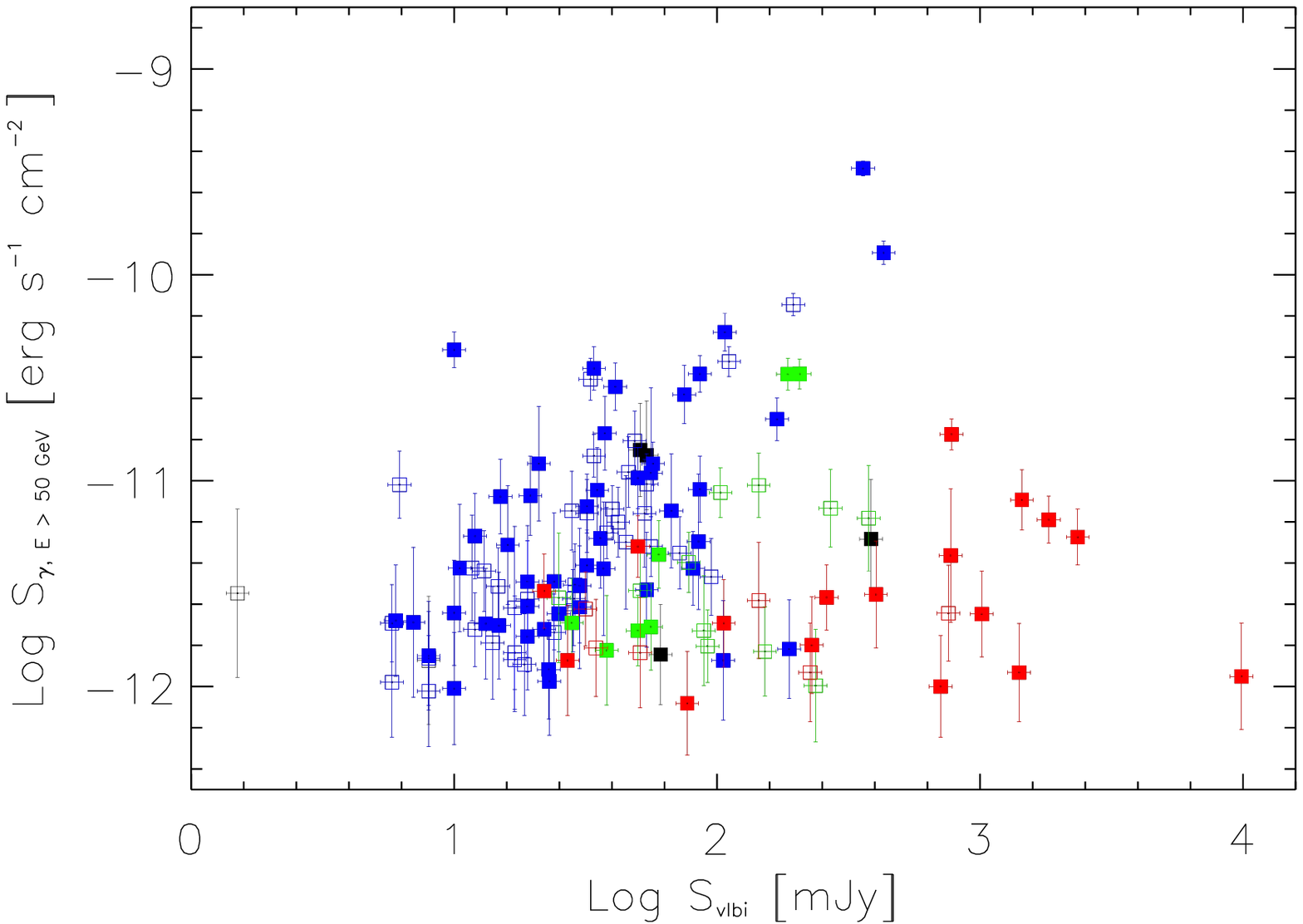} 
\includegraphics[bb=0 0 485 348, width= 0.45\textwidth, clip]{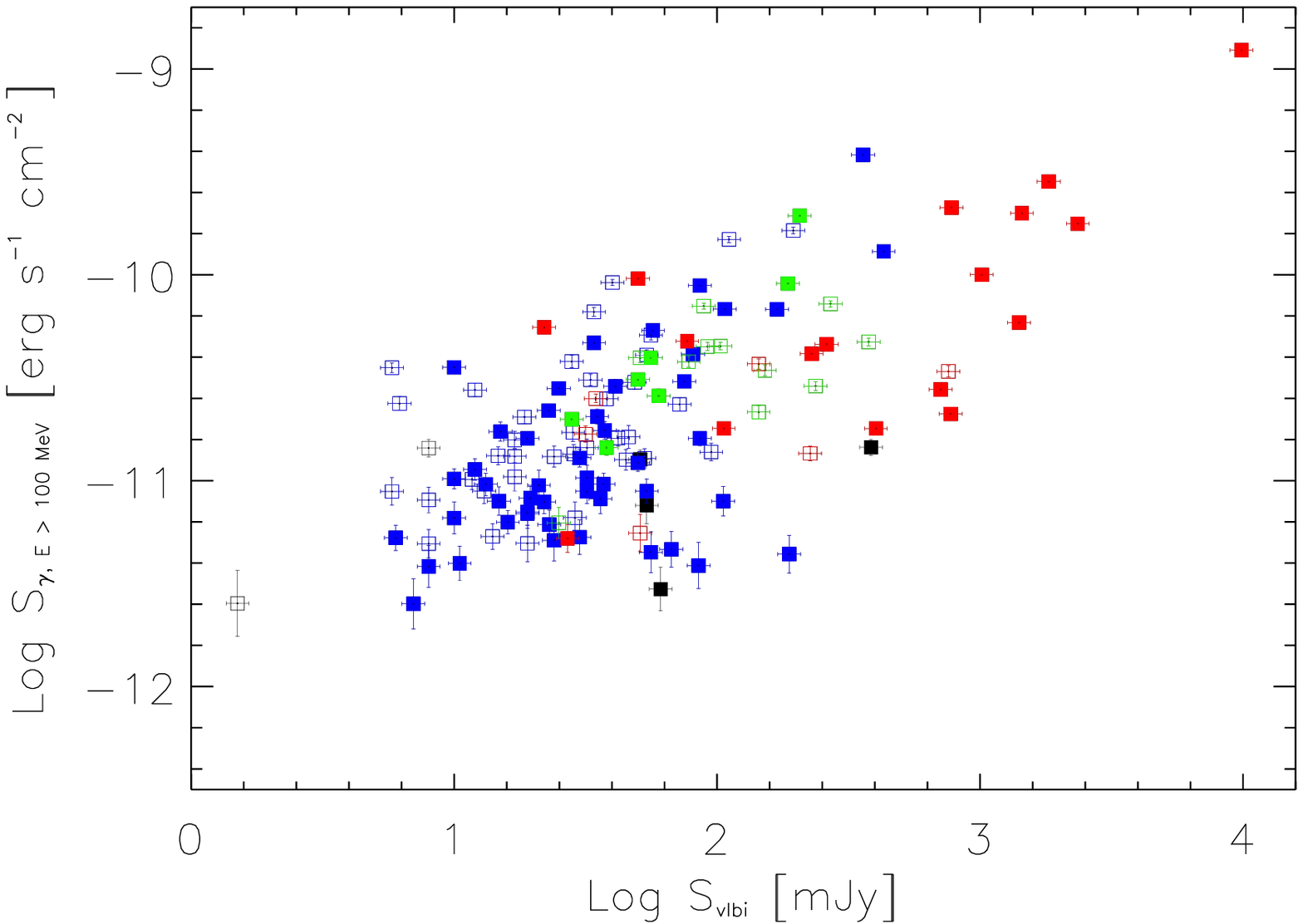}  \\
\end{center}
\caption{\small VLBI flux density vs. 2FHL (left panel) and 3FGL (right panel) energy flux scatter plots for the full 2FHL-n sample. HSP, ISP and LSP sub-classes are indicated in blue, green and red colors, respectively. Sources with no spectral classification are indicated in black color. The filled and empty symbols represent sources with or without redshift, respectively.}
\label{2fhl_scatter_plot_combo}
\end{figure*}

\begin{figure*}
\begin{center} 
\includegraphics[bb=5 155 490 355, clip]{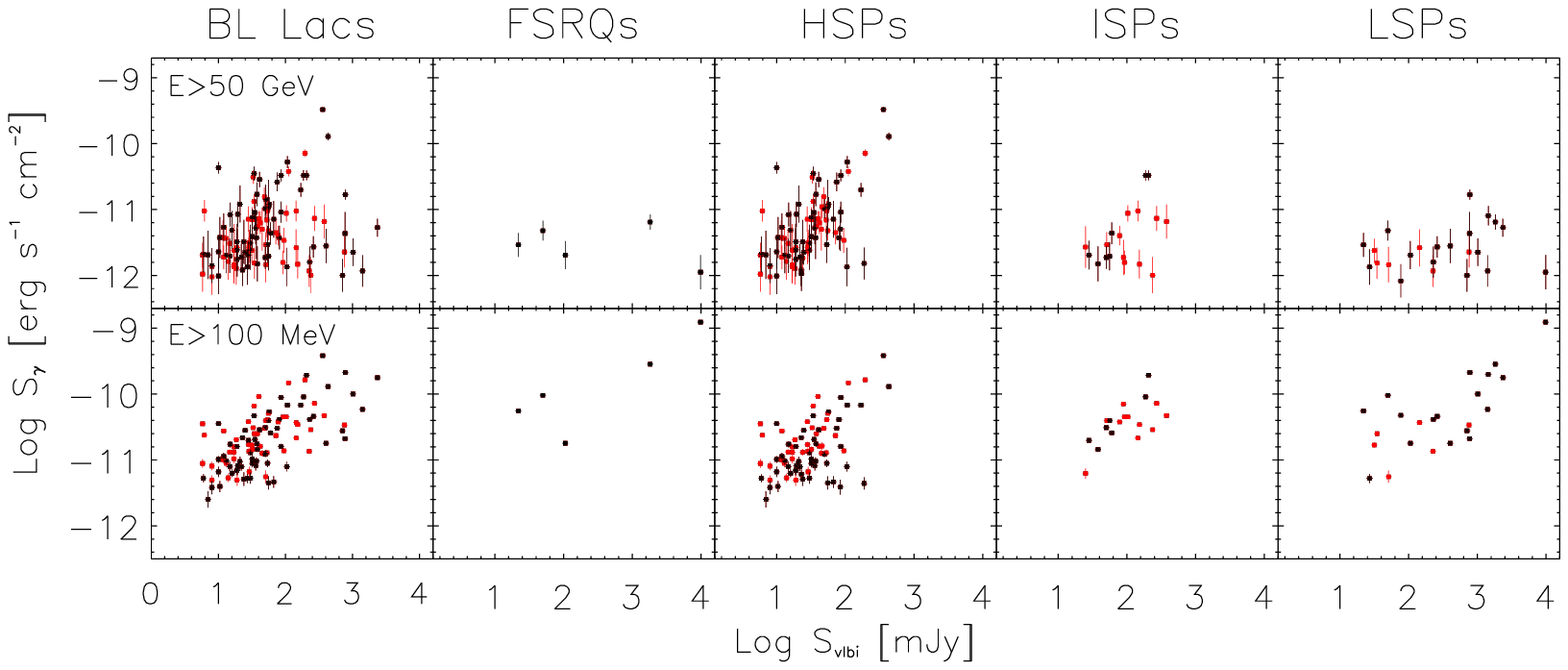} 
\end{center} 
\caption{\small VLBI flux density vs. 2FHL (upper panels) and 3FGL (lower panels) energy flux scatter plots for BL Lacs, FSRQs, HSPs, ISPs,and LSPs, belonging to the 2FHL-n sample. The black and red symbols represent sources with or without redshift, respectively.}
\label{2fhl_scatter_plots}
\end{figure*}

\begin{table*}[!h]
\centering
\caption{\small Results of the correlation analysis between 2FHL (50\,GeV - 2\,TeV) energy fluxes and VLBI flux densities for various 2FHL-n sub-samples. For comparison the same analysis was performed by using 3FGL (0.1 - 300\,GeV) energy fluxes.}
\label{tab_2fhl_corr}     
\small
\setlength{\tabcolsep}{9pt}
\renewcommand{\arraystretch}{1.3}              
\begin{tabular}{lccccc}        
\hline
\hline  
Source type & Catalog & Number of Sources & Number of $z$-bins & r-Pearson & Significance \\  
\hline                        
All sources & 2FHL & 76  & 7  &  0.13 & $ 0.36 $              \\
            & 3FGL & 76  & 7  &  0.72 & $ <10^{-6} $          \\
BL Lac      & 2FHL & 63  & 6  &  0.23 & $ 0.34 $              \\
            & 3FGL & 63  & 6  &  0.73 & $ <10^{-6} $          \\
HSP - with $z$& 2FHL & 48  & 4  &  0.57 & $ 7.0\times 10^{-6} $ \\
              & 3FGL & 48  & 4  &  0.58 & $ <10^{-6} $          \\
HSP - all${\scriptsize ^1}$ & 2FHL & 84  & 8  &  0.61 & $ <10^{-6} $          \\
            & 3FGL & 84  & 8  &  0.53 & $ <10^{-6} $          \\
\hline
\end{tabular}
\begin{threeparttable}
\begin{tablenotes}
\item [1] \footnotesize Full 2FHL-n HSP sample. For those sources without known $z$, we assign a redshift randomly selected among the 2FHL-n source sample with known redshifts (see Sect.~\ref{2fhl-n_analysis}). 
\end{tablenotes}
\end{threeparttable}
\end{table*}

\section{Results}
\label{sec.results}
In Sect.~\ref{1fhl-n_analysis} we investigate the correlation between radio VLBI and $\gamma$-ray emission at $E > 10$\,GeV for the 147 sources with known $z$ of the 1FHL-n sample. By using the same strategy, in Sect.~\ref{2fhl-n_analysis} we perform the correlation analysis by considering the 76 2FHL-n sources with known $z$ detected in the energy range between 50\,GeV and 2\,TeV.

We explore the possible correlation for the full sample and for the different subsets of blazars, divided according to the optical (BL Lacs and FSRQs) and spectral (HSPs, ISPs and LSPs) classification, by using
1FHL/2FHL energy fluxes and the 3FGL energy fluxes as a reference. 

To assess the statistical significance of the correlation results between radio VLBI and $\gamma$-ray emission we use the method of surrogate data proposed by \citet{Pavlidou2012}, that was used in the analysis presented in \citet{Ackermann2011}.
This method, based on permutations of the luminosities, takes into account the various observational biases (e.g.\,, Malmquist bias and common distance effects) which can apparently enhance or dilute any intrinsic luminosity correlation.
Since the method of surrogate data requires the calculation of luminosities, for the correlation analysis we only consider sources with known redshift. As a consequence the number of sources in the samples could be significantly reduced and the redshift distribution may be altered. This condition mainly affects the class of BL Lacs and HSP objects: only about half of them, in the 1FHL-n and 2FHL-n samples, have redshifts.
\citet{Pavlidou2012} and \citet{Ackermann2011} showed that the correlation significance generally increases when more sources are added, for reasonable assumptions on the redshift distribution of sources without a known $z$.

\subsection{1FHL-n AGN sample}
\label{1fhl-n_analysis}
The distribution of the $\gamma$-ray energy fluxes above 10\,GeV of the 1FHL-n sample ($S_{\gamma, \mathrm{1FHL}}$) is shown in the top-left panel of Fig.~\ref{histo_flux} (solid black line). $S_{\gamma, \mathrm{1FHL}}$ has a median value of $\sim 6.4 \times 10^{-12}$ \eflux and covers about three orders of magnitude, ranging from $1.1 \times 10^{-12}$ to $3.2 \times 10^{-10}$ \eflux.
BL Lacs and FSRQs are represented by blue and red dashed line, respectively, with BL Lacs reaching the highest $\gamma$-ray energy flux values.
The VLBI flux density distribution is shown in the top-right panel of Fig.~\ref{histo_flux} (solid black line), and has a median value of $57$ mJy. BL Lacs tend to cluster at lower flux density values (median value $42$ mJy) than FSRQs (median value $372$ mJy).

In Fig.~\ref{1fhl_scatter_plot_combo} we show the scatter plots of the VLBI flux density vs. 1FHL (left panel) and 3FGL (right panel) energy flux. The different colors represent the three spectral types: LSP (red), ISP (green) and HSP (blue) objects. The filled and empty symbols indicate those sources with known and unknown redshift, respectively.
In Fig.~\ref{1fhl_scatter_plots} we show the scatter plots of the VLBI flux density vs. 1FHL (upper panel) and 3FGL (lower panel) energy flux for each blazar sub-class (BL Lacs, FSRQs, HSPs, ISPs, and LSPs).
The results of the correlation analysis are summarized in Table~\ref{tab_1fhl_corr}. We report the number of sources in each subset, the number of redshift bins used for the permutations (with each bin containing at least 10 objects), the resulting Pearson product-moment correlation coefficient ($r$), and the corresponding statistical significance ($p$), which represents the probability to obtain a correlation, from intrinsically uncorrelated data, at least as strong as the one observed in the real sample.

When we use the 3FGL energy fluxes (0.1 - 300\,GeV) we find a strong positive correlation with a high statistical significance (chance probability $p<10^{-6}$) for the full sample ($r = 0.71$), and for all of the considered HSP/ISP/LSP and FSRQ/BL Lac blazar sub-classes (see Table~\ref{tab_1fhl_corr}).

On the other hand, by considering the full 1FHL-n sample, VLBI flux densities and 1FHL energy fluxes are uncorrelated ($r = -0.05$). Even when we separately consider BL Lacs and FSRQs we do not find any correlation between radio VLBI and $\gamma$-ray at $E > 10$\,GeV emission. 
The correlation coefficient shows a different behavior when each spectral blazar sub-class is considered.
HSPs are the only blazar sub-class showing a strong ($r = 0.57$) and significant ($p = 1\times10^{-6}$) correlation, while for LSP and ISP objects in the 1FHL energy range we find a weak correlation.

HSP blazars are therefore the only blazar sub-class showing a strong and significant correlation between radio VLBI and 1FHL/3FGL $\gamma$-ray emission.

\subsection{2FHL-n AGN sample}
\label{2fhl-n_analysis}
In the bottom-left panel of Fig.~\ref{histo_flux} we show the distribution of the $\gamma$-ray energy fluxes above 50\,GeV ($S_{\gamma, \mathrm{2FHL}}$) of the 2FHL-n sample (solid black line).
$S_{\gamma, \mathrm{2FHL}}$ ranges from $8.3 \times 10^{-13}$ to $3.3 \times 10^{-10}$ \eflux, with a median value of $\sim 3.2 \times 10^{-12}$ \eflux.
BL Lacs and FSRQs are represented by blue and red dashed line, respectively.
We show the VLBI flux density distribution in the bottom-right panel of Fig.~\ref{histo_flux} (solid black line), which has a median value of $41$ mJy. BL Lacs and FSRQs have median values of $38$ and $106$ mJy, respectively.

In Fig.~\ref{2fhl_scatter_plot_combo} we show the scatter plots of the VLBI flux density vs. 2FHL (left panel) and 3FGL (right panel) energy flux. The different colors represent the three spectral types: LSP (red), ISP (green) and HSP (blue) objects. The filled and empty symbols indicate those sources with known and unknown redshift, respectively.
The scatter plots of the VLBI flux density vs. 2FHL (upper panel) and 3FGL (lower panel) energy flux for each blazar sub-class (BL Lacs, FSRQs, HSPs, ISPs, and LSPs) are shown in Fig.~\ref{2fhl_scatter_plots}.

We note that in the 2FHL-n sample for some blazar sub-classes the number of sources with known $z$ is less that 20 (e.g.\,, there are only five objects classified as FSRQs) and therefore is not large enough for obtaining a statistically significant result. For this reason, we perform the correlation analysis only for the full 2FHL-n sample, for the BL Lac class, and for the HSP sub-sample.
In Table~\ref{tab_2fhl_corr} we summarize the correlation analysis results by reporting the number of sources in each subset, the number of redshift bins used for the permutations (with each bin containing at least 10 objects), the correlation coefficient $r$, and the corresponding statistical significance.

When the 3FGL energy fluxes are used we find a strong correlation for all the considered blazar sub-classes. On the other hand, VLBI flux densities and 2FHL energy fluxes are uncorrelated both for the full sample and for BL Lac objects (see Table~\ref{tab_2fhl_corr}). On the contrary, for blazars of HSP type a strong ($r = 0.57$) and significant ($p = 7\times10^{-6}$) correlation is found.

HSP objects are again the only blazar sub-class for which we find a strong and significant correlation both in the 2FHL and 3FGL $\gamma$-ray energy ranges.

As mentioned earlier, the method of surrogate data for the statistical significance can only be applied to sources with known $z$.
This requirement can play an important role in the case of HSP objects of our sample, considering that for about half of them the redshift is unknown.
For this reason, we perform the correlation analysis for the full 2FHL-n HSP sample (84 objects), by assigning a redshift value for the sources without known redshift, randomly selected from the sources of the 2FHL-n sample with known redshifts. In this way we assume the same redshift distribution.
We find a correlation coefficient $r = 0.61$, with $p <10^{-6}$ (see Table~\ref{tab_2fhl_corr}).
We note that similar results are obtained if we assume that the unknown redshifts are systematically higher ($\Delta z=0.5$) than the known ones.

\section{Discussion and conclusions}
\label{sec.discussion}
As revealed by the observations performed by \fermi -LAT and the ground-based IACTs, blazars dominate the census of the $\gamma$-ray sky. Exploring the possible correlation between radio and $\gamma$-ray emission is a central issue to understand the blazar emission processes and physics.

A strong and significant correlation between radio and $\gamma$-ray emission was found in several works \citep[e.g.\,,][]{Kovalev2009, Ghirlanda2010, Nieppola2011}. However, the correlation strength seems to decrease when higher $\gamma$-ray energies are considered \citep[see ][]{Ackermann2011, Mufakharov2015}.
In the present work we explore the possible existence of a correlation between radio and GeV-TeV $\gamma$-ray emission, by using the most complete and unbiased AGN samples available at present, extracted from the 1FHL and 2FHL catalogs.
An important novelty of our analysis is that at radio frequencies we use VLBI flux densities, which are more representative of the innermost source regions, where the $\gamma$-ray emission is produced, than single dish or interferometric observations with arcsecond-scale resolution.

The present work points out that (1) HSP blazars are the only sub-class showing a strong ($r = 0.6$) and significant ($p <10^{-6}$) correlation between radio VLBI emission and $\gamma$ rays with $E > 10$\,GeV, (2) the radio-$\gamma$-ray correlation is found for all classes when the 0.1 - 300\,GeV 3FGL energy range is considered.

The correlation that we find when we consider the 0.1 - 300\,GeV LAT $\gamma$-ray band is stronger than what was found by \citet{Ackermann2011}. This result may be a direct consequence of the fact that at radio frequencies we use VLBI flux densities, which probe the radio emission from the regions close to the $\gamma$-ray emission zone, while previous works considered low-resolution radio data with possible contamination from extended structures.
Such a strong correlation between radio and $\gamma$-ray at $E > 100$\,MeV emission was also revealed by \citet{Ghirlanda2011} in a sample of 230 \fermi\ AGNs. In that work the authors made use of 20\,GHz Australia telescope compact array observations, such a high frequency is representative of the emission from the jet base, with no significant contribution from the large-scale structures.

The strong and significant radio and $\gamma$-ray connection vanishes when $\gamma$ rays approaching the VHE domain are considered for all of the blazar sub-classes, with the exception of HSP objects. 
This effect, suggested in previous dedicated analysis, is well constrained and quantified in the present work by using the largest AGN samples currently available at $E > 10$\,GeV and by taking into account the various observational bias.

\begin{figure}
\begin{center}
\includegraphics[bb= 84 74 690 550, width= 0.45\textwidth, clip]{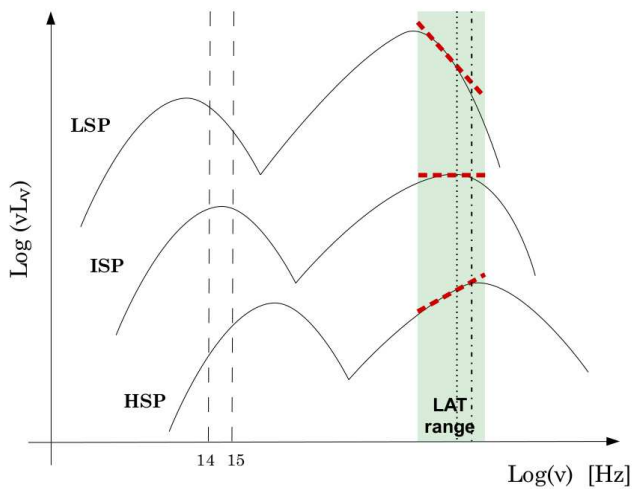} 
\end{center}
\caption{\small Schematic representation of LSP (upper curve), ISP (middle curve) and HSP (lower curve) blazar spectral classification, according to the position of $\nu^{\mathrm{Syn}}_{\mathrm{peak}}$. The green filled area represents the $0.1 - 300$\,GeV 3FGL energy range, while the black dotted and dash-dotted vertical lines represent the 1FHL and 2FHL energy thresholds, respectively.}
\label{schema_SEDs}
\end{figure}

We explain the results of the present analysis within the framework of the blazar SED and its interpretation.
In Fig.~\ref{schema_SEDs} we show a schematic representation of the LSP (upper curve), ISP (middle curve) and HSP (lower curve) blazar spectral classification, according to the position of $\nu^{\mathrm{Syn}}_{\mathrm{peak}}$. The green filled area represents the $0.1 - 300$\,GeV 3FGL energy range.

By inspecting the SED properties of LSP objects it emerges that in general they have soft $\gamma$-ray spectra (with a 3FGL median photon index $\Gamma_{\rm LSP, 3FGL}=2.2$) and their high energy component peak occurs at energies lower than those sampled by the LAT (upper curve in Fig.~\ref{schema_SEDs}).
Moreover, the 1FHL and 2FHL energy ranges (black dotted and dash-dotted vertical lines in Fig.~\ref{schema_SEDs}, respectively) are limited to the highest energies of LAT, where the emission has severely dropped. In the 1FHL and 2FHL energy ranges the median value for the LSP photon index is $\Gamma_{\rm LSP, 1FHL}=2.9$ and $\Gamma_{\rm LSP, 2FHL}=4.3$, respectively, pinpointing a severe steepening of their spectra, and therefore a fast decrease of their flux in the 1FHL/2FHL energy range.
In general the brightest blazars are of LSP type, and their spectral break, above few (1-10)\,GeV, is due to severe cooling losses of the emitting particles \citep[e.g.\,,][]{Tavecchio2009, Finke2010, Harris2012, Stern2014}. In addition, FSRQs are rich of ambient photons that may cause $\gamma \gamma$ absorption, producing an additional energy cutoff.
ISP blazars share most of these features, except that their high energy emission peak may fall in the softest part of the LAT energy band.

HSP objects are less powerful than LSPs/ISPs and the energy losses are less severe. This is directly reflected in the position of the high energy component, which peaks above $\sim100$\,GeV, at much higher energies than LSP/ISP objects. 
In HSP blazars, the part of the high energy spectrum affected by cooling effects is well beyond the energy range sampled by the LAT \citep{Ghisellini1998}, showing a rising spectrum both in the 3FGL and 1FHL/2FHL energy ranges (lower curve in Fig.~\ref{schema_SEDs}). 
This is reflected in their harder spectra than those of LSPs ($\Gamma_{\rm HSP, 3FGL}=1.9$), mostly in the highest $\gamma$-ray energy ranges ($\Gamma_{\rm HSP, 1FHL}=2.1$ and $\Gamma_{\rm HSP, 2FHL}=2.8$). 
The connection of the observed bolometric luminosity and the shape of the blazar SED is described by the so-called blazar sequence, in which both the low- and high-energy emission SED peaks shift to lower frequencies when the total power increases \citep{Fossati1998, Ghisellini1998, Ghisellini2017}. 
As a consequence, for the brightest objects, in general of LSP type, in the 1FHL/2FHL energy range we are sampling the part of the spectrum where the high energy emission is strongly decreasing. On the contrary, for HSP objects the high energy emission SED peak is usually found within 1FHL/2FHL energy range. 
This sampling effect, which mainly affects LSP objects, can be connected with the fact that we find a correlation between radio and gamma-ray emission only for HSPs.


Regarding the optical blazar sub-classes, we note that when the $0.1 - 300$\,GeV 3FGL energy range is considered a strong correlation is found for both optical blazar sub-classes, with BL Lacs showing a higher correlation coefficient ($r = 0.70$) with respect to FSRQs ($r = 0.49$). The different correlation strength may be ascribed to the intrinsically different properties of the two optical blazar sub-classes. 
The rich ambient photon field and the usually higher distance of FSRQs with respect to BL Lacs make their $\gamma$-ray spectrum softer, likely weakening the correlation. 
However, when $\gamma$ rays at $E > 10$\,GeV are considered the correlation vanishes for both FSRQs and BL Lacs. This is because the FSRQ and BL Lac classification is only based on the properties of their optical spectra without taking into account the different energy and spectral properties.

The sources of both 1FHL-n and 2FHL-n samples span a wide range of $z$ (from 0.01 up to 2.2), and the redshift distribution is different among the different blazar sub-classes.
For this reason, to further validate our results we run the correlation analysis by using K-corrected radio flux densities and $\gamma$-ray energy fluxes. When using K-corrected quantities it is important to have a reliable estimation of both $z$ and the spectral indexes in the two observing bands for not introducing additional uncertainties. By assuming an average spectral index $\alpha = 0$ in the radio band, and by using the best-fit power-law photon index provided by the 3FGL catalog in the $\gamma$-ray energy band, we obtain results which are in good agreement and consistent with those presented in Tables~\ref{tab_1fhl_corr} and \ref{tab_2fhl_corr}.

Within this simple picture, an important issue to be taken into account is the variability argument.
Blazars are strongly variable objects at all frequencies, showing intensity variations on timescales ranging from several years to a few days. In particular, at TeV energies they show variability on timescales as short as a few minutes \citep[e.g.\,, ][]{Aharonian2007}. 
By considering that our data are not taken simultaneously and that we are using average values for the $\gamma$-ray fluxes and radio flux densities from single observations, the variability can affect and spoil a possible correlation for these sources.
In particular, being the variability more pronounced above the SED peak, this effect is more relevant for LSP and ISP objects, whose high energy emission SED peak occurs at lower energies than those sampled in the 1FHL/2FHL energy range. 
Conversely, given that in HSP objects the high energy emission SED peak is found in general at energies above $\sim100$\,GeV, in the 1FHL/2FHL energy range they are not as variable as LSPs/ISPs, and their correlation should be less affected by the use of non simultaneous data.

\citet{Ackermann2011} revealed for the first time that the correlation between radio and $\gamma$-ray emission is stronger when concurrent observations are considered. 
However, also in the case of concurrent observations there are some caveats to be taken into account. The radio and $\gamma$-ray emission vary on different time scales, with the $\gamma$-ray variability being in general more rapid. Blazars often show strong outbursts or long-term periods of enhanced activity both at radio and $\gamma$-ray frequencies observed with time lags, due to the optical depth effects at radio frequencies \citep[see e.g.\,,][]{Ghirlanda2011, Fuhrmann2014}. 

As it emerges from the results of the present analysis, to proper characterize the radio vs. VHE emission connection, both extensive long-term VLBI monitoring and systematic VHE sky surveys are required. The new generation aperture synthesis radio telescope Square Kilometer Array (SKA) in synergy with the new generation ground-based VHE $\gamma$-ray instrument CTA will provide us with the best chance to investigate the existence of radio-VHE emission connection.

\begin{acknowledgement}
\begin{small}
We acknowledge financial contribution from grant PRIN-INAF-2014. 
For this paper we made use of NASA's Astrophysics Data System and the TOPCAT software \citep{Taylor2005}.
This research has made use of the NASA/IPAC Extragalactic Database NED, which is operated by the JPL, Californian Institute of Technology, under contract with the National Aeronautics and Space Administration. 
The \textit{Fermi}-LAT Collaboration acknowledges support for LAT development, operation, and data analysis from NASA and DOE (United States), CEA/Irfu and IN2P3/CNRS (France), ASI and INFN (Italy), MEXT,
KEK, and JAXA (Japan), and the K.A.~Wallenberg Foundation, the Swedish Research Council, and the National Space Board (Sweden). Science analysis support in the operations phase from INAF (Italy) and CNES (France) is also gratefully acknowledged.
\end{small}
\end{acknowledgement}

\end{document}